\documentclass[epj]{svjour}
\usepackage{amsmath}
\usepackage{amsbsy}
\usepackage{MnSymbol}
\usepackage{graphicx}
\usepackage{subfigure}
\usepackage{fancyref}

\begin{document}

\title{Velocity autocorrelation function of a Brownian particle} 
\author{D. Chakraborty
  \inst{1} \thanks{chakraborty@itp.uni-leipzig.de}}
\institute{Institute for Theoretical Physics, University of Leipzig,
  \\Vor dem Hospitaltore 1, 04103 Leipzig, Germany}

\abstract{ In this article, we present molecular dynamics study of the
  velocity autocorrelation function (VACF) of a Brownian particle. We compare
  the results of the simulation with the exact analytic predictions
  for a compressible fluid from \cite{Chow1973} and an approximate
  result combining the predictions from hydrodynamics at short and
  long times. The physical quantities which determine the decay were
  determined from separate bulk simulations of the Lennard-Jones fluid
  at the same thermodynamic state point.We observe that the long-time
  regime of the VACF compares well the predictions from the
  macroscopic hydrodynamics, but the intermediate decay is sensitive
  to the viscoelastic nature of the solvent.  }
\maketitle

\section{Introduction}
Brownian motion is the random motion of a particle, which is large
compared to the solvent molecules, but is not of macroscopic size.  It
has become a paradigm in various branches of science and remains an
active area of research among theoreticians and experimentalists. It
is not only a preferred tool of theoretical modeling, but is also
extensively used to probe microscopic environments in
experiments\cite{Wirtz2009,Squires2010,Frey2005}. A considerable
effort is also spend in investigating transport properties of
colloidal suspensions and complex fluids,primarily due to the
relevance of such systems in industry, using molecular simulations of
Brownian motion
\cite{Vladkov2006,Keblinski2002,Padding2006,Lee2004,Eastman2004}.
Such simulations always invariably involve using discrete particles to
explain continuum predictions of theory, and therefore, requires a
clear understanding of the molecular and continuum regimes.

In this article, using molecular dynamics simulation, we investigate
the velocity autocorrelation function (VACF) of a Brownian
particle. We choose a large system size so that the effect of
finite-size of the simulation box is small. The internal degrees of
freedom of the nanoparticle is also resolved in the simulation, such
that stick boundary conditions apply on the surface of the
particle\cite{Li2009}. The erratic motion of a Brownian particle
exhibits a far more rich behavior than predicted by the simple
Langevin picture.  In the continuum description, a fluid is well
described by the Stokes equation, with the particle dynamics coupled
to the solvent through the imposed boundary conditions. The inadequacy
of the Langevin picture in describing such erratic motion can be
immediately seen from the VACF of the Brownian particle.  While a
simple exponential decay is predicted at all times in the Langevin
model, in reality the decay exhibits distinct features of both
continuum, as well as discrete nature of the solvent.  Accordingly, we
classify the decay in three separate regimes, a short-time regimes -
where molecular nature of the solvent plays a crucial role, an
intermediate regime - governed by the interplay between sound
propagation, vorticity diffusion and the viscoelasticity of the
solvent, and a long-time regime where the VACF decays as a power law
$t^{-d/2}$ ($d$ is the dimension of space) due to the development of
the slow viscous patterns in the solvent
\cite{Alder1967,Alder1970,Hauge1973,Chow1973,Herschkowitz-Kaufman1972}.
We compare the results from the simulations with the exact predictions
from hydrodynamics, and observe, that, while the decay of the VACF at
long times compares well with the predictions from hydrodynamics, the
intermediate decay is sensitive to the viscoelastic nature of the
fluid and can not be explained by only considering the compressible
nature of the fluid.

Using a molecular dynamics simulation to investigate short and
long-time dynamics of isothermal Brownian motion is a non-trivial
task. The key points in such simulations, are the identification of
relevant length and time scales. The two important length
scales in the system are the simulation box size $L$ and the radius of
the particle $R$. In a typical molecular dynamics simulation periodic
boundary conditions are imposed, implying that the dynamics of the
particle is effected by its periodic images. Since the strength of
such finite size effect is determined by the ratio of the two length
scales, $R/L$, we have the choice of a small $R$ or large $L$. Both
of these choices are unfortunately restricted. While the choice of $L$ is
solely determined by the computational resources at hand, the choice
of $R$ is determined by a number of factors. Ideally, one would prefer
a clear separation of the time scales in the simulation, in
particular, the sonic time $\tau_c=R/c$ and the vorticity diffusion
time $\tau_\nu = R^2/\nu$, both of which determine the decay of the
VACF. A small choice of $R$ does not resolve these time scales
properly.

The lower bound for $R$ is determined by the Knudsen number, defined
as the ratio of the mean free path to the characteristic length scale
of the flow -- typically, the diameter of the particle. The
Knudsen number decides whether a continuum or a statistical mechanics
description of the system is appropriate. Besides the length scales,
the time scales involved in a Brownian motion range from the order of
$10^{-15}\,$s to seconds. The simulation must also be able to resolve
the various time scales in the problem, the smallest of which is the
collision time of solvent molecules and the largest time scale is the
colloid diffusion time, over which the colloid diffuses over its own
radius.

The remainder of the article is organized as follows.  In \Fref{sec:md_sim},
we explain our molecular dynamics simulation in brief. The comparison
of the simulation results with the exact prediction from the theory is
done in \Fref{sec:vacf_exact}, and an approximate result is presented
in \Fref{sec:vacf_app}.

\section{Molecular Dynamics  Simulation}
\label{sec:md_sim}
In this section, we provide the details of our molecular dynamics
simulation. To begin with, the natural choice of units is the
Lennard--Jones reduced units, where length, time and energy in units
of $\sigma$, $\tau=\sqrt{m\sigma^2/\epsilon}$ and
$\epsilon$. Throughout the article, the numerical values of the
physical quantities are given in reduced units, unless otherwise
explicitly stated.

Our model system is made of a simple Brownian particle, with internal
degress of freedom resolved, immersed in a Lennard--Jones solvent.
The particles in the system interact via the Lennard--Jones
interaction, 
\begin{equation}
\label{lj_pot}
U(r)=4\epsilon\left[\left(\frac{\sigma}{r}\right)^{12}-\left(\frac{\sigma}{r}\right)^{6}\right].
\end{equation}
 Additionally, the nearest-neighbor interaction of the atoms in the
spherical cluster of the nanoparticle is the FENE interaction,
\begin{equation}
\label{fene_pot}
U_{FENE}(r)=-\frac{1}{2} \kappa R_0^2 \log \biggr[1-\biggr(\frac{r}{R_0}\biggr)^2 \biggr],
\end{equation}
where $\kappa =30 \epsilon/\sigma^2$ is the spring constant and
$R_0=1.5 \sigma$. In order to keep the finite-size effects from the
image particles to a reasonable value, we choose a large system size
with a total of $256000$ particles. The initial configuration of the
system was chosen to be a perfect FCC lattice with velocities drawn from a
Boltzmann distribution. The nanoparticle was obtained from a spherical
cut of the FCC lattice. For the smallest size of the nanoparticle
($R=3$) there were $177$ atoms in the cluster while for largest size
of the nanoparticle ($R=5$) we have $767$ atoms in the cluster. The
ratio $L/R$, which quantifies the finite-size in the system are
$22.5795\,$ and $13.5409\,$ for the $R=3$ and $R=5$, respectively.

The system was first equilibrated under NPT ensemble, with the system
coupled to a thermostat and barostat, at a thermodynamic pressure of
$P_0=0.01$ and temperature $T_0=0.75$.  For the implementation of the
NPT ensemble we chose the N\'ose-Hoover equations of motion as
modified by Melchionna\cite{Melchionna1993},
\begin{eqnarray}
\label{4}
\nonumber
\mathbf{\dot{r}}_i &=& \mathbf{\dot{v}}_i + \alpha (\mathbf{\dot{r}}_i - \mathbf{R}_0), \phantom{1} \mathbf{\dot{p}}_i = \mathbf{F}_i - (\alpha + \gamma) 
(\mathbf{\dot{r}}_i - \mathbf{R}_0)\\
\nonumber
\dot{\gamma} &=& \nu_T^2 \biggr( \frac{T(t)}{T_0} -1\biggr),\phantom{1} \dot{\alpha} = \frac{\nu_p^2}{N k_B T_0} V(p(t)-p),\\
\dot{V}&=& 3 V \alpha .\\
\nonumber
\end{eqnarray}

The barostating variable $\alpha$ can be eliminated between the last
two equations in Eq.(\ref{4}), and the resulting equations are then
numerically integrated using Leap-Frog integration
scheme\cite{Toxvaerd1993}.

The molecular dynamics simulations were implemented on Graphics
Processing Units (GPU) and is similar to those of Anderson
et. al. \cite{Anderson2008} and Colberg \cite{Colberg2009}, more closely
resembling the later in the construction of the Verlet list. We briefly
describe our implementation in the following lines.

We use the atom decomposition method in the simulation, for efficient
parallel implementation of our MD code. Every particle in the
simulation is assigned a thread, which is responsible for updating the
coordinates and momenta of the particle. A GPU optimized cell list
algorithm is used for construction of the Verlet list. For this
purpose, the simulation domain is divided into cubes of size $r_c$,
where $r_c$ is the cutoff length scale for the Lennard-Jones potential
($r_c=2.5 \sigma$ in the present application). The particles were
first sorted into their respective cells using the parallel radixsort
algorithm \cite{Colberg2009}. To construct of the Verlet list, the
entries of $26+1$ cells are copied to the shared memory. Every cell
has an upper limit for the maximum number of entries, determined by
the size of the shared memory on the GPU. This limitation also
prevented us from simultaneously copying the particle coordinates to
the shared memory.  For a given particle in a cell, an iterative
search is made of the neighboring cells and the particle coordinates
are read from a texture array. The stability and numerical accuracy of
our molecular dynamics code was verified by outputting the total
energy and total momentum of the system. With an integration time-step
$\delta t=0.001$, trajectories of $2 \times 10^7$ steps (corresponding
to a physical duration of $20\,$ ns) were simulated for the data points.

\section{Velocity Autocorrelation Function}
\label{sec:vacf_exact}
The long time tails in the VACF can be explained by the generalized
Langevin equation,
\begin{equation}
\label{gen_lan_eq}
\dot{\mathbf{P}}=-\frac{1}{M}\int_0^t \zeta(t-t')\mathbf{P}(t')+\boldsymbol{\xi}(t)
\end{equation}
together with the time dependent friction coefficient $\zeta(t)$,
where $\mathbf{P}$ is the momentum of the Brownian particle and
$\boldsymbol{\xi}$ is the random force acting on it. Multiplying
Eq.(\ref{gen_lan_eq}) by $\mathbf{P}(0)$, and taking an ensemble
average, the velocity autocorrelation function for the Brownian
particle becomes,
\begin{equation}
\label{vacf_time}
\dot{C}(t)=\frac{1}{M}\int_0^t \zeta(t-t')C(t'),
\end{equation}
which, in the frequency space is written
as,
\begin{equation}
\label{vacf_omega}
\tilde{C}(\omega)=C(0)\left(-i\omega
  +\frac{\tilde{\zeta}(\omega)}{M}\right)^{-1} \equiv C(0) Y(\omega).
\end{equation}
The zero-frequency limit of Eq.(\ref{vacf_omega})
produces the Stokes-Einstein relation $D=k_{\rm B}T/\zeta_0$.
Transforming back to real time, the normalized VACF of the Brownian
particle is given by
\begin{equation}
  \label{norm_vacf}
  \frac{C(t)}{C(0)}=\int_{-\infty}^{\infty}\frac{\mathrm{d}\omega}{2\pi}Re[Y(\omega)]\cos(\omega t)
\end{equation}

For an incompressible fluid, the frequency dependent
friction coefficient is given by \cite{Hauge1973,Zwanzig1970}:
\begin{equation}
\label{zeta_incomp}
\tilde{\zeta}(\omega)= 6 \pi \eta_0 R(1 +R(-i\omega/\nu)^{1/2} -(i\omega
R^2/9\nu)), 
\end{equation}
where, $\eta_0$ and $\nu$ are the steady-state dynamic and kinematic
viscosities of the solvent,respectively. The square-root singularity
in Eq.(\ref{zeta_incomp}) gives rise to the power law decay of the
VACF at long times. Because of the incompressibility condition, the
equal-time value of the VACF suffers a discontinuity from the
equipartition value of $k_B T/M$ to $k_B T/M^{*}$, with the effective
mass $M^{*}$ given by $M^{*}=M+M_f/2$. $M_f$ is the mass of the
displaced fluid.

In simulations, however, the discontinuity is not observed due to the
finite compressibility of the fluid
\cite{Zwanzig1970,Bakker2002,Chow1973}. In a compressible solvent, the
sound propagation occurs with a finite speed, and the solvent
surrounding the colloid is not instantly set in motion. A fraction of
the energy of the Brownian particle is thus spent in creating these
sound waves. To account for the compressibility effect of the solvent,
we need to consider the Boussinesq force for unsteady motion in a
compressible solvent\cite{Zwanzig1970,Chow1973}. For this purpose, we
consider the frequency dependent friction coefficient presented in the
works of Chow et. al. \cite{Chow1973}.  For simplicity, we follow the
notations in \cite{Chow1973}. In terms of the vorticity diffusion time
$\tau_\nu=R^2/\nu$ and the sonic time $\tau_c=R/c$,
$\tilde{\zeta}(\omega)$ is written as
\begin{eqnarray}
\label{zeta_chow}
\nonumber
&x&=i\sqrt{i \omega \tau_\nu}, \,\,\,\,\, y=i\omega \tau_c\left[1-\frac{i
  \omega \tau_c^2}{\tau_\nu}\dfrac{(\mu+(4/3)\eta)}{\eta}\right]^{-1/2}\\
&\tilde{\zeta}(\omega)&=-\frac{4}{3}\pi \eta R x^2 [(1-y)Q+2(x-1)P],
\end{eqnarray}
where $P$ and $Q$  are functions of the dimensionless
variables $x$ and $y$. We refer the readers to \cite{Chow1973}, for
an explicit expression of these functions. Substituting
Eq.(\ref{zeta_chow}) in Eq.(\ref{vacf_omega}), and subsequently using
Eq.(\ref{norm_vacf}), we obtain the normalized VACF in real time. 

The physical parameters which enter Eq.(\ref{zeta_chow}) were
determined from separate molecular dynamics simulations of the bulk
Lennard-Jones fluid at the same thermodynamic state point. The shear
viscosity $\eta$ and the bulk viscosity $\mu$ were estimated from the
off-diagonal and diagonal components of the stress tensor and using
the Green-Kubo formula,
\begin{equation}
  \label{eta_t}
  \eta(t)=\frac{V}{k_B T} \int_0^t \langle \sigma_{xy}(t')\sigma_{xy}(0) \rangle \mathrm{d} t'
\end{equation}
\begin{equation}
\label{mu_t}
\mu(t)=\frac{V}{k_B T} \int_0^t \langle \delta p(t') \delta p(0) \rangle \mathrm{d} t'
\end{equation}
The infintie time limit of Eq.(\ref{eta_t}) and Eq.(\ref{mu_t})
provided the steady state values of the shear and bulf viscosity,
$\eta_0$ and $\mu_0$, respectively. The adiabatic sound speed in the
solvent was estimated from the relation
\begin{equation}
\label{sound_speed}
c^2=\frac{\gamma}{\rho m \chi_T},
\end{equation}
where $\gamma$ is the ratio of the specific heats $C_P/C_V$, $\rho$ is
the density of the solvent, $m$ is the mass of the solvent particles
and $\chi_T$ is the isothermal thermal compressibility.

\begin{figure}
\includegraphics[width=\linewidth]{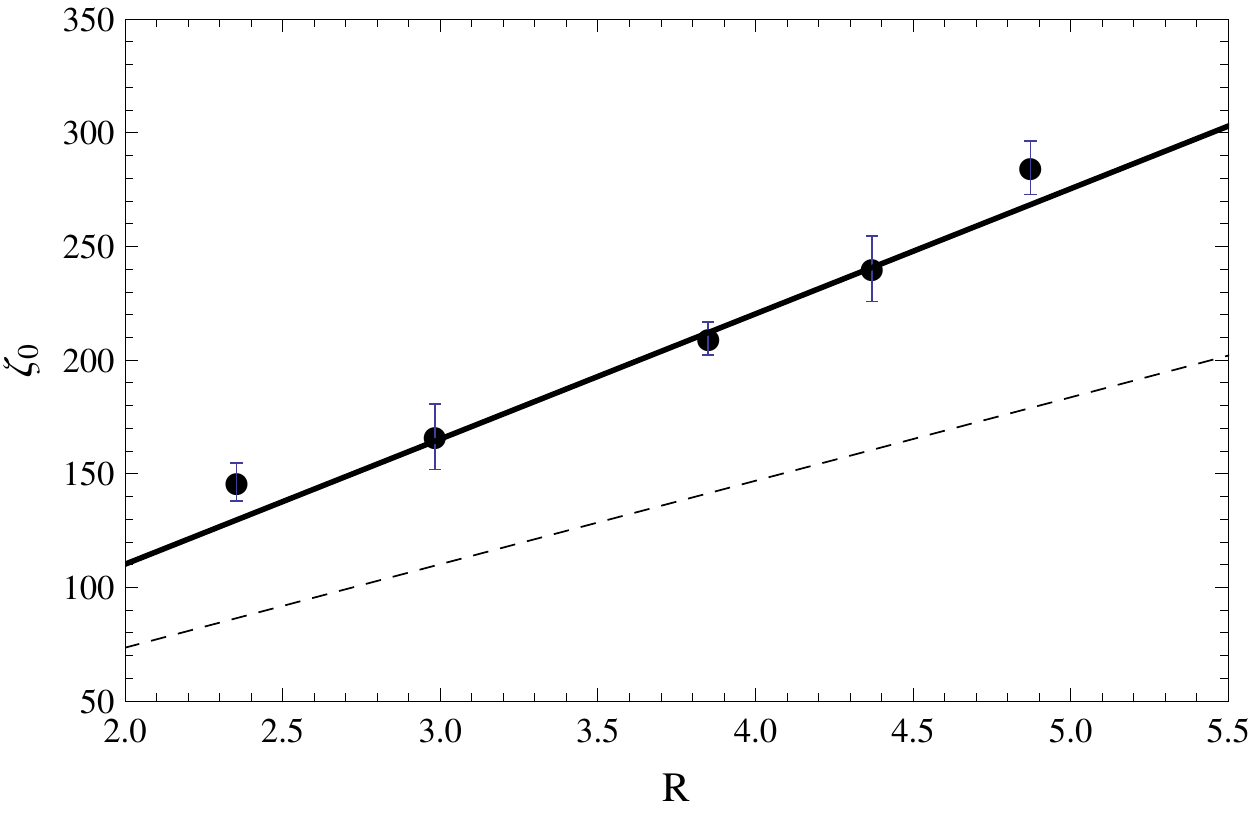}
\caption{The dependence of the friction coefficient for different
  radius of the nanoparticle. The solid line is plot of Stokes law for
  stick boundary condition, $\zeta_0=6 \pi \eta_0 R$, while the dashed
  line shows the plot of $\zeta_0 = 4 \pi \eta_0 R$. The radius of the
  nanoparticle were estimated from the Eq.(\ref{radius}).}
\label{fric}
\end{figure}

Before we compare the results of the molecular dynamics simulations
with the predictions from hydrodynamics, the assumptions of the
macroscopic theory needs to validated. The crucial assumption which
enters the theory is the boundary condition on the surface of the
particle. While Eq.(\ref{zeta_chow}) assumes a stick boundary
condition on the particle surface, such an assumption may break down
in a microscopic scale. To validate this we measured the friction
coefficient for different radius of a rough Brownian particle. In
figure \ref{fric}, we show this dependence of the friction coefficient
on the radius of the Brownian particle and compare it with predictions
from Stokes law with stick and slip boundary condition. The reasonable
agreement of the steady-state friction coefficient with the Stokes law
$\zeta_0=6 \pi \eta_0 R$ indicates that hydrodynamic boundary
conditions applicable on the surface of the sphere are those of stick
boundary conditions.\footnote{When the radius of the Brownian particle
  is comparable to the size of the solvent particles, the
  Stokes-Einstein relation with standard stick or slip boundary
  conditions can break down and a non-standard boundary condition may
  be required \cite{Li2009}. However, for the particle sizes for which
  the velocity autocorrelation function has been investigated in the
  present article, stick boundary conditions are valid as depicted in
  Figure \ref{fric}.}
Additionally, the Kundsen number (Kn) for the system was determined by
measuring the mean-free path of the solvent from the ballistic regime
of the solvent mean-square displacement. The Knudsen number determines
whether a statistical mechanics or a continuum description is more
appropriate for the system and for large values of Kn (typically Kn
$>0.1$), deviations from the continuum description become
relevant. The measured values of Kn were $\mathrm{Kn} = 0.02261$ and
$\mathrm{Kn}=0.01366$, for $R=3$ and $R=5$, respectively, which
indicates that the assumptions in the macroscopic hydrodynamics
remains valid in the present scenario.

\begin{figure}
\includegraphics[width=\linewidth]{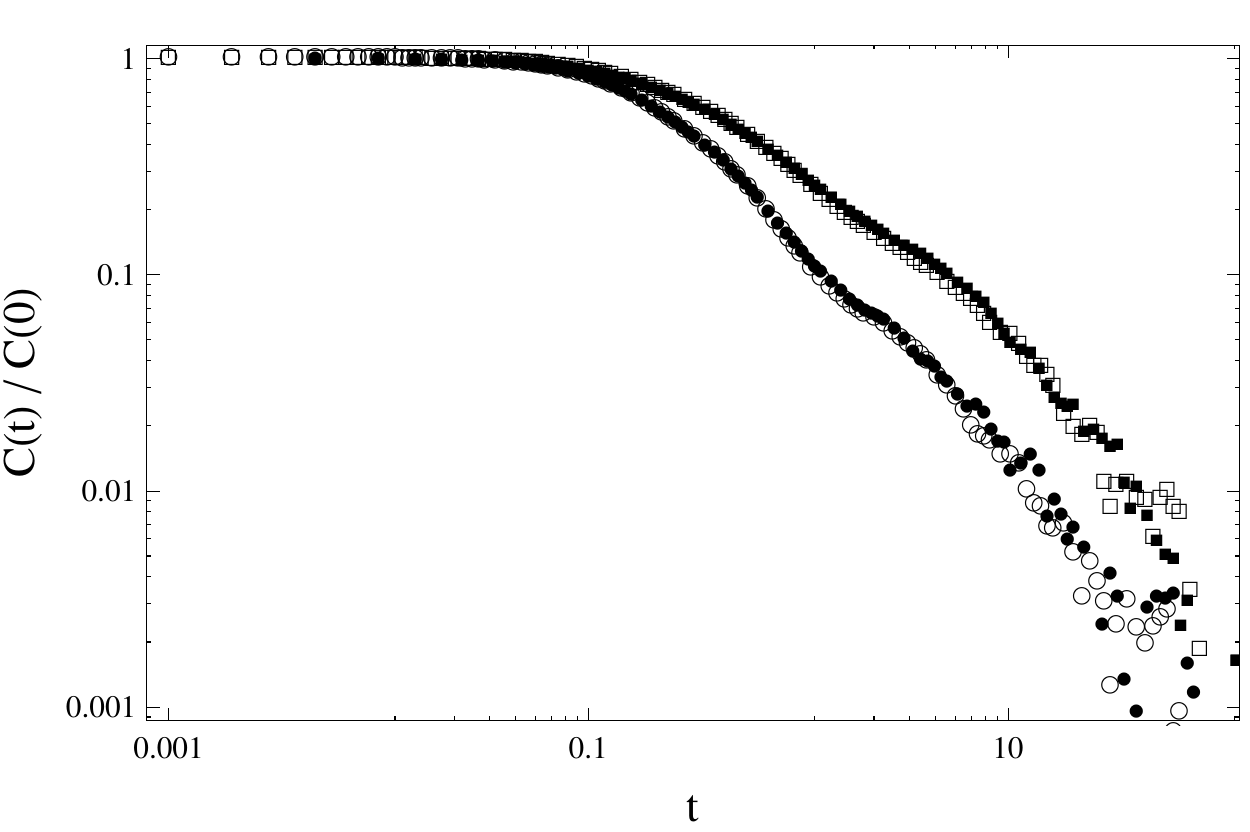}
\caption{Velocity autocorrelation function of a Brownian particle with
  radius $R=5$ $(\square, \blacksquare)$ and $R=3$ ({\Large $ \circ,
    \bullet$}) for two different sizes of the simulation box,
  $L\approx 68$ ($\square$,{\Large $\circ$}) and $L \approx 85$
  ($\square$,{\Large $\bullet$}). }
\label{vacffntsz}
\end{figure}

Moreover, due to the periodic boundary condition imposed in the
simulations, the data suffers from finite-size effects from the
long-ranged flow field of the image particles. We chose a sufficiently
large simulation box, so that the artifact of the image particles are
small. To substantiate this, we performed simualtions with two
different box lengths, $L\approx 68$ and $L \approx 87$, the result of
which is depicted in the Figure \ref{vacffntsz}. The measured velocity
autocorrelation functions does not exhibit pronounced finite-size
effects, particulary in the intermediate regime of interest.

\begin{figure}
\includegraphics[width=\linewidth]{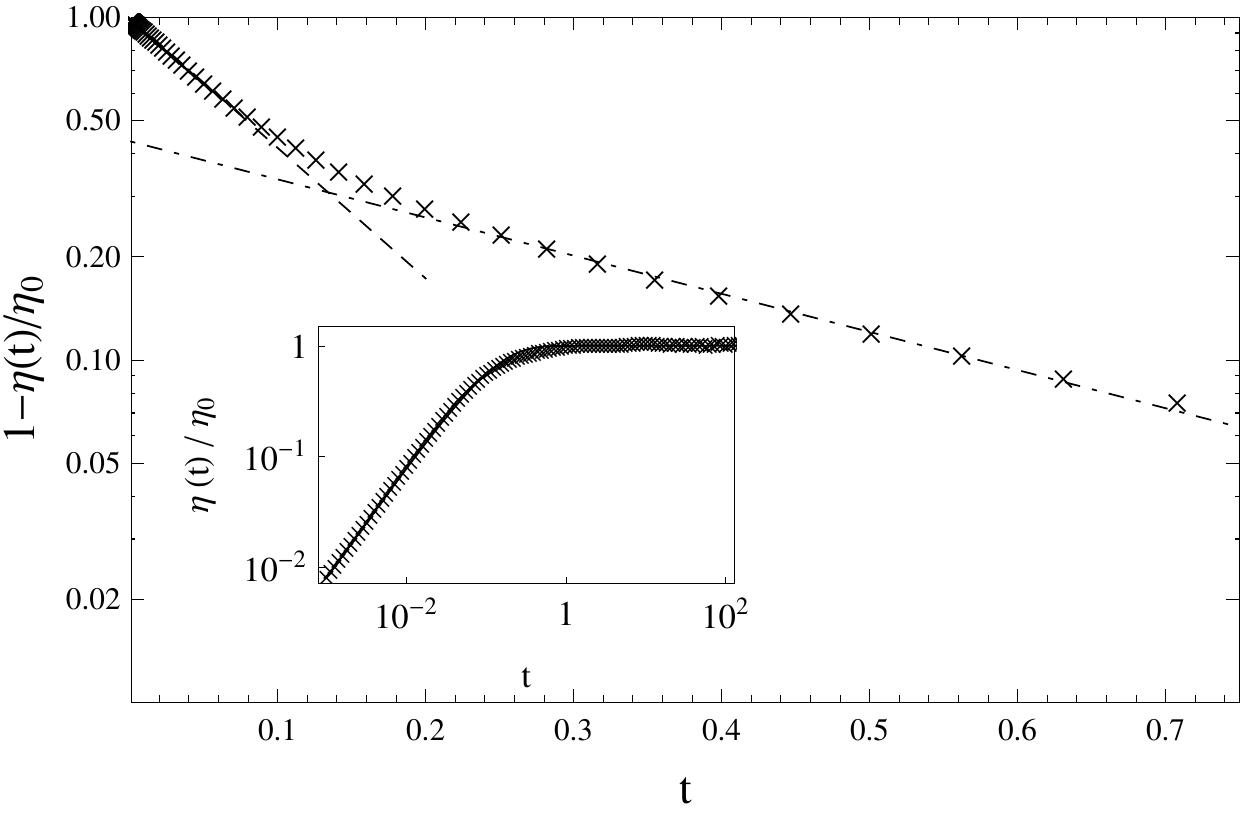}
\caption{Decay of $1-\eta(t)/\eta_{0}$ with time. Two distinct
  decay-time scales are seen. At short times, comparable with
  molecular collision time, the decay is faster compared to the
  intermediate regime where power-law decay of the stress
  autocorrelation function becomes important. In the time-scales of
  interest, the decay can be modeled as simple exponential with decay
  time $\tau_1$ and $\tau_2$. The two time constants $\tau_1$
  (corresponding to the dashed line) and $\tau_2$ (corresponding to
  the dot-dashed line) obtained from the above plot are $0.114$ and
  $0.390$, respectively. The inset shows the time dependent viscosity
  normalized by $\eta_0$.}
\label{eta_decay}
\end{figure}

To compare the reuslts from the simulation with the theoretical
predictions for the VACF, the numerical evaluation of
Eq.(\ref{norm_vacf}) was first carried out with the steady state
values of the shear viscosity. However, we observed that the
intermediate decay of the VACF can not be accurately described by only
treating the solvent as compressible and it was essential to consider the
viscoelastic nature of the solvent \cite{Zwanzig1970}. As already
pointed out in \cite{Grimm2011}, the interaction of a colloid in a
viscoelastic solvent can be visualized by considering the colloid
connected to the fluid by a dash-pot and spring in series. At times
larger than $\tau_\nu$, the viscous dissipation is represented by the
dash-pot, while at shorter times the colloid interacts with the fluid
via elastic forces. This caging effect is more pronounced when the
ratio of the mass of the Brownian particle to the solvent particles is
small \cite{Ould-Kaddour2000}. To this end, we model the Lennard-Jones
solvent as Maxwell fluid with the frequency dependent viscosity given
by
\begin{equation}
\label{maxwell_fld}
\tilde{\eta}(\omega)=\eta_0/(1-i \omega \tau),
\end{equation}
where $\eta_0$ is the steady-state shear viscosity of the solvent
($\omega=0$ component).  The relaxation time $\tau$ is
related to the infinite frequency shear modulus $G_\infty$ as $\tau
=\eta_0/G_\infty$. For a Lennard-Jones solvent, the single exponential
relaxation in Eq.(\ref{maxwell_fld}) ignores the algebraic decay of
the stress autocorrelation at long times. This is also evident from
the time dependent viscosity obtained from the simulation using
Eq.(\ref{eta_t}). To illustrate this more clearly, we plot the
variation of the quantity $1-\eta(t)/\eta_{0}$ with time in
Fig. \ref{eta_decay}. The two distinct decays shown in Fig. \ref{eta_decay} 
can be modelled as simple exponential with decay times $\tau_1$ and
$\tau_2$.
To determine the VACF in the intermediate regime using
Eq.(\ref{norm_vacf}) and Eq.(\ref{zeta_chow}) , we use $\tau_2$ as the
relaxation time for the model fluid. Since the colloid is made of
discrete number of particles, the radius of the sphere $R$ was determined
from the radius of gyration using the relation
\begin{equation}
  \label{radius}
  \langle R_g^2 \rangle = \frac{1}{N}\sum_i^N (\mathbf{r}_i-\mathbf{R}_{\rm CM})^2 =\frac{3}{5}R^2.
\end{equation}
In the numerical evaluation of the VACF, we observed that the radius
of the particle which gives a more accurate fit to the data was close
to $R+\sigma/2$, where $\sigma$ is the diameter of the Lennard-Jones
fluid particles. The values of $R$ used were $5.35$ and $3.33$
compared to the value of $R+\sigma/2=5.37$ and $R+\sigma/2=3.42$,
respectively. In table \ref{tab:table_A}, we compare the numerical
values of the physical quantities $\eta$, $\mu$ and $c$ for a bulk
Lennard-Jones fluid and those which provided the best fit to the
simulated velocity autocorrelation funtion using Eq.(\ref{norm_vacf}) .
Additionally, the mass of the colloidal particle was always taken as
the reduced mass of the system.
\begin{figure}
  \label{vacf_plt}
\includegraphics[width=\linewidth]{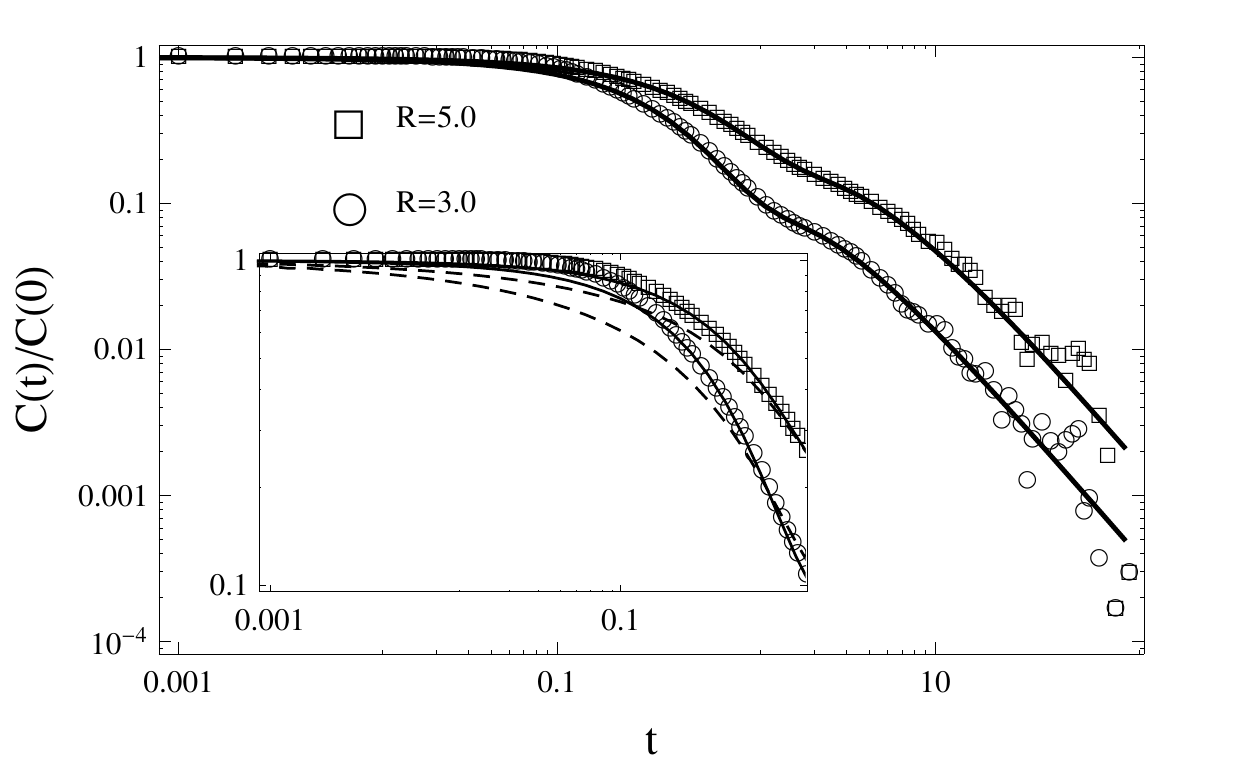}
\caption{Decay of the velocity autocorrelation function of a Brownian
  particle for two particle radius $R=5$ ($\Box$) and $R=3$
  ($\medcircle$). The solid-line is numerical evaluation of the
  normalized VACF using Eq.(\ref{norm_vacf}) and Eq.(\ref{zeta_chow})
  together with a frequency dependent viscosity
  (Eq.(\ref{maxwell_fld})). The decay in the intermediate regime is
  sensitive to the viscoelastic nature of the solvent, and can not be
  accurately described using only the assumption of compressible fluid
  (see inset). The dashed line in the inset is the numerical
  evaluation with a constant viscosity.}
\end{figure}

\begin{table}
\centering
\begin{tabular}{c c c}
Physical & From bulk & Value used in nu-\\
quantities & simulation & merical evaluation\\
\hline
$\eta$ & $2.79961 \pm 0.12963$ & $2.79961$\\
$\mu$ & $0.78556 \pm 0.07442$ & $0.78561$\\
$c$ & $5.18256 \pm 0.05225 $ &  $5.15444$ \\
\end{tabular}
\caption{Comparison of the numerical values of the physical quantities $\eta$, $\mu$ and $c$, obtained from the bulk simulations of Lennard--Jones fluid and from best fit of the simualted VACF using Eq.(\ref{norm_vacf}). We note, that the values of these parameters agree within statistical error bars.}
\label{tab:table_A}
\end{table}

\section{Approximate Result for the VACF}
\label{sec:vacf_app}
The inverse transformation of Eq.(\ref{norm_vacf}) is only possible
numerically and an exact closed form analytical expression for the
VACF is difficult. However, an an approximate result can be formulated
using the argument of time-scale separation between $\tau_c$ and
$\tau_\nu$. The sound waves created in the solvent always precedes the
development of slow viscous patterns\cite{Espanol1995}, with $\tau_c$
usually an order of magnitude larger compared to $\tau_\nu$. Assuming
this time scale separation, the complete decay of the VACF can be
constructed by a simple addition of the VACF at short and long-time
regime \cite{Padding2006}. 

In the long-time regime, following \cite{Hauge1973}\cite{Paul1981},
the normalized VACF of a Browninan particle can be written as:
\begin{equation}
 \label{vacf_hydro}
 M\langle V(t)V(0)\rangle /k_B T= \frac{2 \rho_c}{3 \rho_f}\frac{1}{3 \pi}\int_0^{\infty} \mathrm{d}x \ \frac{e^{-x t/{\tau_\nu}} x^{1/2}}{1+\sigma_1 x +\sigma_2 x^2}
\end{equation}
with $\sigma_1=(1/9)(7-4\rho_c/\rho_f)$ and
$\sigma_2=(1/81)(1+2\rho_c/\rho_f)^2$ . In Eq.(\ref{vacf_hydro}),
$\rho_c$ is the density of the colloid and $\rho_f$ is the density of
the fluid.  At $t=0$, the integral gives the value $\pi
(1-\sqrt{\sigma_1^2-4\sigma_2})/\sqrt{2(\sigma_1-\sqrt{\sigma_1^2-4\sigma_2})}$,
and the right-hand side of Eq.(\ref{vacf_hydro}), after
simplification, becomes $(2\rho_c/\rho_f)/(1+2\rho_c/\rho_f)$,
producing the well known discontinuity. The discontinuity is quite
easily removed when we take into account the finite compressibility of
the fluid. To this end, we consider the general expression of Zwanzig
\cite{Zwanzig1975},
\begin{eqnarray}
  \label{vacf_sound}
  M\langle V(t)V(0)\rangle/k_B T= \frac{e^{-\alpha_1 t/\tau_c}}{(1+2\rho_c/\rho_f)}\bigg[ \text{Cos}\left(\frac{\alpha_2 t}{\tau_c}\right)\\
  \nonumber
  -\frac{\alpha_1}{\alpha_2}\text{Sin}\left(\frac{\alpha_2 t}{\tau_c}\right)\bigg]
\end{eqnarray}
with $\alpha_1=(1+\rho_f/2\rho_c)$ and $\alpha_2=(1-\rho_f^2/4\rho_c^2)^{1/2}$.
\begin{figure}[!t]
  \includegraphics[width=\linewidth]{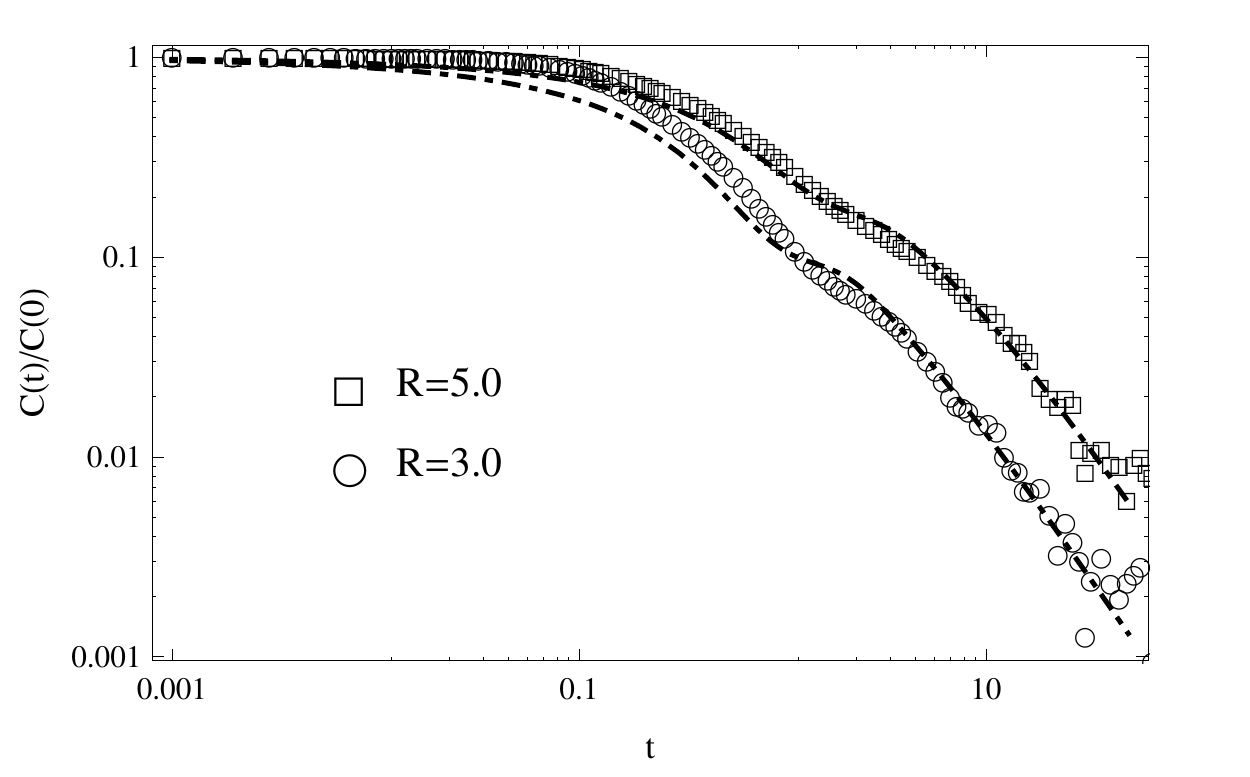}
  \caption{Comparison of the theoretical prediction of VACF with
    simulation for two values of the radius of the particle. The
    dot-dashed lines is the plot of addition of Eq.(\ref{vacf_hydro})
    and Eq.(\ref{vacf_sound}). Even though the fit to the simulation
    data is poor in the short and intermediate regime, it still
    reproduces the qualitative nature of the decay. The physical
    parameters used in the plot are the same as presented in
    \Fref{tab:table_A} }
\label{vacf_app}
\end{figure}
For a neutrally buoyant particle ($\rho_c=\rho_f$) the two
contribution takes the form
\begin{equation}
 \label{vacf_hydro_nb}
\langle V(t)V(0)\rangle =\frac{2 k_B T}{3M}\frac{1}{3 \pi}\int_0^{\infty} \mathrm{d}x \ \frac{e^{-x t/{\tau_\nu}} x^{1/2}}{1+x/3 +x^2/9}
\end{equation}
and
\begin{eqnarray}
  \label{vacf_sound_nb}
  \nonumber
  \langle V(t)V(0)\rangle =\frac{K_B T}{3 M}e^{-(3/2)(t/{\tau_c})}\bigg[\text{Cos}\bigg(\frac{\sqrt{3}t}{2\tau_c}\bigg) \\
  -\sqrt{3}\text{Sin}\bigg(\frac{\sqrt{3}t}{2\tau_c}\bigg)\bigg]
\end{eqnarray}
To a first order, the complete VACF of a colloidal particle can be
described by a simple addition of Eq.(\ref{vacf_hydro_nb}) and
Eq.(\ref{vacf_sound_nb}):
\begin{eqnarray}
 \label{vacf_comp_nb}
 \nonumber
 \langle V(t)V(0)\rangle =\frac{2 k_B T}{3M}\frac{1}{3 \pi}\int_0^{\infty} \mathrm{d}x \ \frac{e^{-x t/{\tau_\nu}} x^{1/2}}{1+x/3 +x^2/9}\\
 \nonumber
 +\frac{K_B T}{3 M}e^{-(3/2)(t/{\tau_c})}\bigg[\text{Cos}\bigg(\frac{\sqrt{3}t}{2\tau_c}\bigg)-\sqrt{3}\text{Sin}\bigg(\frac{\sqrt{3}t}{2\tau_c}\bigg)\bigg]\\
\end{eqnarray}

In Fig.\ref{vacf_app}, we compare the normalized VACF of the Brownian
particle with the theoretical predictions obtained by addition of
Eq.(\ref{vacf_hydro}) and Eq.(\ref{vacf_sound}).

\begin{figure}[!h]
  \includegraphics[width=\linewidth]{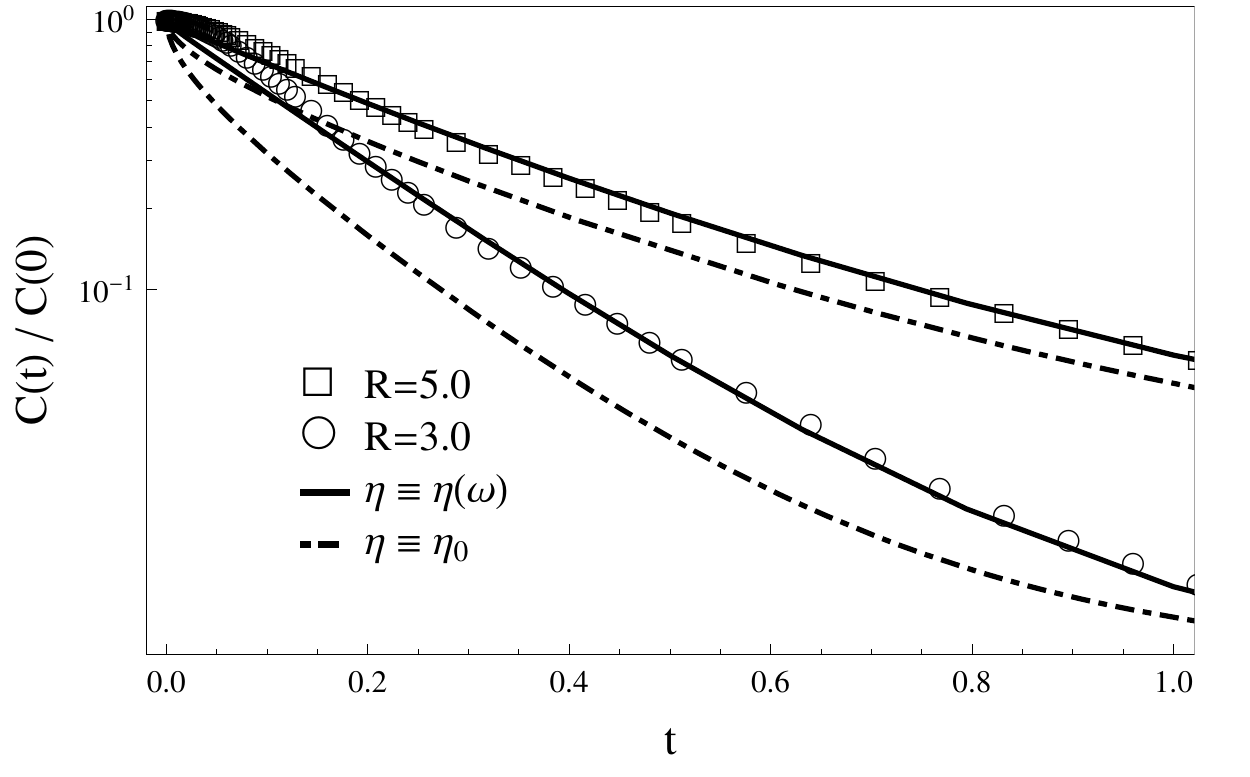}
  \caption{Comparison of the VACF of a Brownian particle in the
    intermediate decay regime. A reasonable fit to the data is
    obtained only when we consider the viscoelastic nature of the
    solvent.  The solid line in the plot is the numerical evaluation
    of the VACF using a frequency dependent viscosity, and the
    dot-dashed lines is the plot of addition of Eq.(\ref{vacf_hydro})
    and Eq.(\ref{vacf_sound}). }
\label{vacf_comp_int}
\end{figure}

Finally, we compare the VACF of a Brownian particle in the
intermediate regime in Fig.\ref{vacf_comp_int}. The intermediate
decay, between the molecular collision time and the sonic time, 
is clearly sensitive to the viscoelastic nature of the fluid. 

In conclusion, using molecular dynamics simulations, we have
investigated the decay of the velocity autocorrelation function (VACF)
of a colloid in a Lennard-Jones solvent. The numerical values of the
shear and kinematic viscosities and the speed of sound, which
determine the decay, were obtained from separate simulations of the
bulk Lennard-Jones fluid at the same thermodynamic state point.  These
values were used to determine VACF from the exact analytical
prediction. Accordingly, we divide the complete decay in three
regimes, a short-time regime where discrete nature of the fluid plays
an important role, an intermediate regime - governed by the interplay
between sound propagation, vorticity diffusion and viscoelasticity of
the fluid and a long-time regime of algebraic decay due to vorticity
diffusion. We observe, that the decay in the intermediate regime can
not be accurately described by only considering the compressibility of
the fluid, but the viscoelastic nature of the solvent should also be
taken into account. 

\section*{Acknowledgments}
The author gratefully acknowledges stimulating discussions with Jens
Glaser (Minnesota) and Klaus Kroy (Leipzig). This work was supported
by Deutsche Forschungsgemeinschaft (DFG) via FOR 877 and the Alexander
Von Humboldt Foundation.

\bibliography{library}
\bibliographystyle{plain}
\end{document}